\documentclass[aps,pra,twocolumn,groupedaddress,10pt]{revtex4-1}

\usepackage{amsmath}
\usepackage{graphicx}
\usepackage{multirow}
\usepackage{bm}
\usepackage{caption}
\usepackage{array}
\usepackage{color}
\usepackage{csquotes}

\begin{document}
\title{Unidirectional Reflectionless Transmission for  Two-Dimensional 
$\mathcal{PT}$-symmetric Periodic  Structures}

\author{Lijun Yuan}
\email{Corresponding author: ljyuan@ctbu.edu.cn}
\affiliation{College of Mathematics and Statistics, Chongqing Technology and Business University, Chongqing, 
China}
\author{Ya Yan Lu}
\affiliation{Department of Mathematics, City University of Hong Kong, Hong Kong}

\begin{abstract}
Unidirectional reflectionless propagation (or transmission) is an
interesting wave phenomenon observed in many $\mathcal{PT}$-symmetric optical
structures. Theoretical studies on unidirectional
reflectionless transmission often use simple  coupled-mode models.
The coupled mode theory can reveal the most important physical mechanism
for this wave phenomenon, but it is only an approximate theory, and it 
does not provide accurate quantitative predictions with respect to
geometric and material parameters of the structure. In this paper, we
rigorously study
unidirectional reflectionless transmission for two-dimensional (2D)
$\mathcal{PT}$-symmetric periodic structures sandwiched between two
homogeneous media. Using a scattering matrix  formalism and a
perturbation method,  we show that
real zero-reflection frequencies are robust under
$\mathcal{PT}$-symmetric perturbations, and unidirectional
reflectionless transmission is guaranteed to occur if the 
perturbation (of the dielectric function) satisfies a simple
condition. Numerical examples are presented to
validate the analytical results, and to demonstrate unidirectional
invisibility by tuning the amplitude of balanced gain and loss. 
\end{abstract}

\maketitle


\section{Introduction}
In recent years, $\mathcal{PT}$-symmetry 
has attracted considerable attention 
in the optics and photonics community \cite{bender98,feng17,longhi17,miri19}. 
A $\mathcal{PT}$-symmetric optical structure is usually realized by a 
complex dielectric function with a symmetric real part and an anti-symmetric imaginary part (i.e.  a balanced gain and loss). 
The $\mathcal{PT}$-symmetry 
provides a fertile and feasible tool to manipulate lightwaves. 
Many interesting wave phenomena have been observed on
$\mathcal{PT}$-symmetric optical structures. Noticeable examples include
unidirectional reflectionless propagation \cite{lin11,longhi11,kalish12,mostaf13,sarisa17,fu16,rivolta16,rege12,feng13,
huang17,horsley15,yang16,sarisa18}, single-mode lasing \cite{miri12,liu17}, and simultaneous lasing and coherent perfect absorption \cite{longhi10,chong11}. 

Unidirectional reflectionless propagation (or transmission) is the  phenomenon wherein the reflection is zero for an incident wave coming from one side and nonzero for an incident wave coming from the other side. 
For lossless dielectric structures with certain symmetry,  it is well known that zero reflection and zero transmission can really occur \cite{popov86,shipman12,chesnel18}, but unidirectional reflectionless transmission is impossible, because the unitarity of the scattering matrix implies that 
zero reflections for left and right incident waves must occur at the same frequency. 
This is not the case for  $\mathcal{PT}$-symmetric structures, since
the scattering matrix is no longer unitary \cite{chong11,ge12}. A
particularly interesting case of unidirectional reflectionless
transmission is unidirectional invisibility, for which the transmitted
wave is identical to that without the local structure \cite{lin11}. 
Unidirectional reflectionless transmission has been studied on  various $\mathcal{PT}$-symmetric optical structures, including one-dimensional  (1D) structures \cite{lin11,longhi11}, 
planar layered structures
\cite{kalish12,mostaf13,sarisa17,yang16,sarisa18}, 
planar inhomogeneous structure \cite{horsley15}, 
two-dimensional (2D) closed waveguides \cite{fu16},  and 2D coupled waveguide resonator systems \cite{rivolta16}. 
Experimental demonstrations 
have been reported  for $\mathcal{PT}$-symmetric photonic lattices \cite{rege12} and  microscale SOI waveguides \cite{feng13}.  
It should be mentioned that unidirectional reflectionless transmission can also occur in non-$\mathcal{PT}$-symmetric optical structures \cite{feng14,huang15,shen14,gu17,zhao19}.  
Existing studies on unidirectional reflectionless transmission
typically employ 1D Helmholtz equations or  coupled-mode 
models. 

In this paper, we consider 2D $\mathcal{PT}$-symmetric 
periodic structures  sandwiched between two homogeneous
media, and find exact conditions under which 
unidirectional reflectionless 
transmission is guaranteed to occur.
 More specifically, assuming  the  $\mathcal{PT}$-symmetric structure
 is a small  perturbation of a lossless dielectric structure and the
 dielectric structure has a simple (i.e. nondegenerate) real zero-reflection frequency, we show
 that real zero-reflection frequencies continue to exist, and they are 
different for left and right incident waves if the perturbation
satisfies a simple  condition. The continual existence
(i.e. robustness) of real zero-reflection frequencies is proved using
properties of the scattering matrix. A perturbation method is used to
estimate the real zero-reflection frequencies and show that 
unidirectional reflectionless 
transmission occurs at arbitrarily small perturbations. 
 Numerical examples are presented to validate our analytical results,
 and show that unidirectional invisibility can be obtained by tuning
 the amplitude of balanced gain and loss.  

The rest of this paper is organized  as follows. In Sec.~\ref{sec:Smatrix}, we recall some properties of the scattering matrix
for general structures  and discuss zero reflections for
lossless dielectric structures with different symmetries. 
In Sec.~{\ref{sec:PTsym}}, we show that real zero-reflection
frequencies are robust under $\mathcal{PT}$-symmetric perturbations. 
In Sec.~{\ref{sec:perturbation}}, we use a perturbation method to estimate the shifts of the zero-reflection
frequencies for left and right incident waves and derive a condition to
guarantee unidirectional reflectionless transmission. 
In Sec.~{\ref{sec:Numerical}},  
numerical examples are presented to illustrate unidirectional
reflectionless transmission and 
unidirectional invisibility.

\section{Scattering matrix}
\label{sec:Smatrix}
We consider  two-dimensional (2D) structures that are invariant in $z$,  periodic in $y$ with period $L$,  bounded in the $x$ direction, and surrounded by vacuum, where $\left\{ x, y, z \right\}$ is a Cartesian coordinate system. The dielectric function for such a structure and the surrounding media satisfies   
\begin{equation*}
\label{eq:period_eps} \epsilon(x,y+L) = \epsilon(x,y)
\end{equation*}
for all $(x,y)$ and $\epsilon(x,y) = 1$ for $|x| > D$, where $D$ is a given constant.  For the $E$-polarization, the $z$-component of the electric field,
denoted by $u$, satisfies the following 2D Helmholtz
equation:
\begin{equation}
\label{eq:helm}
\frac{\partial^2 u}{\partial x^2} + \frac{\partial^2 u}{\partial y^2} +
k_0^2 \epsilon\, u =  0,
\end{equation}
where $k_0=\omega/c$ is the free space wavenumber, $\omega$ is the angular frequency, and $c$ is the speed of light in vacuum.

In the left and right homogeneous media, we specify  two incident plane waves 
\begin{equation}
\label{eq:incident_left} u^{(i)}_l(x,y) = a_l e^{i[ \alpha (x + D) + \beta y]} \quad \mbox{for} \quad x < -D
\end{equation}
and
\begin{equation}
\label{eq:incident_right} u^{(i)}_r(x,y) = a_r e^{ - i [ \alpha (x -D) - \beta y]} \quad \mbox{for} \quad x > D,
\end{equation} 
 where $a_l$ and $a_r$ are the amplitudes of the incident waves, $(\pm \alpha, \beta)$ are the incident wave vectors, $\beta$ is real, and $\alpha^2 + \beta^2 = k_0^2$.  
Since the  structure is periodic and the medium is homogeneous for $|x| > D$, the solution of Eq.~(\ref{eq:helm}) can be written as 
\begin{equation}
\label{eq:scat_solution_xlessD} u(x,y)=  a_l e^{i[ \alpha (x + D) + \beta y]}+  \sum\limits_{j=-\infty}^{+\infty} b_{lj} e^{- i [ \alpha_j (x + D) - \beta_j y]}, 
\end{equation}
for $x < - D$ and
\begin{equation}
\label{eq:scat_solution_xlargeD} u(x,y) =  a_r e^{ - i [ \alpha (x -D) - \beta y]} +  \sum\limits_{j=-\infty}^{+\infty} b_{rj} e^{i [ \alpha_j (x - D) + \beta_j y]}, 
\end{equation}
for $x > D$,  where $\{b_{lj} \}$ and $\{ b_{rj} \}$ are the amplitudes of the out-going plane waves, and
\begin{equation}
\label{eq:alpha_beta}  \beta_j = \beta + 2 j \pi/ L, \quad \alpha_j = \sqrt{k_0^2 - \beta_j^2} ,
 \end{equation} 
for $ j = 0, \pm 1, \pm 2, \ldots$.
Notice that $\alpha_0 = \alpha$ and $\beta_0 = \beta$.

If  $\beta$ is real,  $\beta \in \left[-\pi/L, \pi/L \right]$,  $\omega$ is real, and $k_0 $   satisfies
\begin{equation}
\label{eq:one_order} | \beta | < k_0 < 2 \pi / L - |\beta|, 
\end{equation}
then $\alpha_0$ is real and all $\alpha_j$ for $j \neq 0$ are pure imaginary with positive imaginary parts. In that case, all out-going plane waves for $j \neq 0$  decay to zero exponentially as $|x| \to \infty$, and the only  out-going propagating plane waves are those for $j \neq 0$, i.e. the plane waves in Eqs.~(\ref{eq:scat_solution_xlessD}) and (\ref{eq:scat_solution_xlargeD}) with coefficients $b_{l0}$ and $b_{r0}$.

Let $S = S(\omega, \beta)$ be the $2 \times 2$ scattering matrix satisfying
\begin{equation}
\label{eq:Smatrix} \left[ \begin{matrix} b_{l0} \\ b_{r0}    \end{matrix} \right] = S(\omega, \beta) \left[ \begin{matrix} a_{l} \\ a_{r}    \end{matrix} \right] = \left[ \begin{matrix} r_L(\omega, \beta) & t_R(\omega, \beta)  \\ t_L(\omega, \beta) & r_R(\omega, \beta)   \end{matrix} \right]  \left[ \begin{matrix} a_{l} \\ a_{r}    \end{matrix} \right],
\end{equation}
where $r_L$ and $r_R$ ($t_L$ and $t_R$) are the reflection (transmission) coefficients for left and right incident waves, respectively,  and $R_L =  |r_L|^2$,   $R_R = |r_R|^2$, $ T_L =  |t_L|^2$ and  $T_R =  |t_R|^2$ are the corresponding reflectance and  transmittance.

Although $\omega$ is generally real, it is possible to study the diffraction problem for a complex frequency. If $\omega$ is allowed to have a small (positive or negative) imaginary part, and the real part of $k_0 = \omega / c$ satisfies Eq.~({\ref{eq:one_order}}), then $k_0^2 - \beta^2$ is close to the positive real axis, and $k_0^2 - \beta_j^2$ (for $j \neq 0$) are close to the negative real axis. In order to define $\alpha_j =\sqrt{k_0^2 - \beta_j^2}$ that depends continuously on the imaginary part of $\omega$, we can use a complex square root with a branch cut on the negative imaginary axis. That is, if $\eta = |\eta| e^{i \psi}$ for $-\pi/2 < \psi \leq 3 \pi /2 $, then $\sqrt{\eta} = \sqrt{|\eta| } e^{i \psi/2}$. Using this square root, Eqs.~(\ref{eq:scat_solution_xlessD}) and (\ref{eq:scat_solution_xlargeD}) are still valid for $|x| > D$, the out-going waves are still dominated by plane waves with coefficients $b_{l0}$ and $b_{r0}$, and the scattering matrix can be defined as in Eq.~(\ref{eq:Smatrix}).

If for a real frequency $\omega$ and a real wavenumber $\beta$, we have $r_L = 0$ and $r_R \neq 0$ (or $r_L \neq 0$ and $r_R = 0$), then  we say unidirectional reflectionless transmission occurs at $(\omega, \beta)$ for a left (or right) incident wave.

The scattering matrix has some important properties. The reciprocity gives rise to
\begin{equation}
\label{eq:Smatrix_reciprocal} S(\omega, -\beta) = S^{\textsf{T}}(\omega, \beta),
\end{equation}
where the superscript ``{\small \textsf{T}}" denotes matrix transpose, and $S(\omega, -\beta)$ is the scattering matrix for incident waves with $e^{- i \beta y} $ dependence. Equation~(\ref{eq:Smatrix_reciprocal}) is a general result valid for complex dielectric function $\epsilon$ and complex $\omega$. For easy reference, we give a proof  in Appendix A. Equation~(\ref{eq:Smatrix_reciprocal}) gives rise to $r_L(\omega, \beta) =  r_L(\omega, -\beta)$, $r_R(\omega, \beta) =  r_R(\omega, -\beta)$ and $t_L(\omega,-\beta) = t_R(\omega,\beta)$. It is clear that if unidirectional reflectionless transmission occurs at a pair $(\omega, \beta)$, then it also occurs at $(\omega, -\beta)$.  If $\beta = 0$, i.e. for normal incident waves, then $t_L = t_R$ for any $\omega$. 

For structures with some symmetries, the scattering matrix can be simplified. 
If the structure has an inversion symmetry, i.e. 
\begin{equation*}
\label{eq:eps_rotationeven} \epsilon(x,y) = \epsilon(-x,-y) \quad \mbox{for all} \quad (x,y),
\end{equation*}
then the mapping $(x,y) \to (-x,-y)$ changes $\beta $ to $-\beta$, swaps $a_r$ with $a_l$ and $b_{l0}$ with $b_{r0}$. This leads to
$$   \left[ \begin{matrix} {b}_{r0} \\ {b}_{l0}    \end{matrix} \right] = S(\omega, -\beta) \left[ \begin{matrix} {a}_{r} \\ {a}_{l}    \end{matrix} \right].$$
Therefore,
\begin{equation*}
 S(\omega, \beta) = P S(\omega, -\beta) P,
\end{equation*}
where 
$$ P = \left[ \begin{matrix}  0 & 1  \\ 1 & 0\end{matrix} \right]. $$
From Eq.~(\ref{eq:Smatrix_reciprocal}), we obtain
\begin{equation}
\label{eq:Smatrix_yeven} S(\omega, \beta) = P S^{\textsf{T}}(\omega, \beta) P.
\end{equation}
This implies that $r_L= r_R$ for all $\omega$ and $\beta$. Therefore, unidirectional reflectionless transmission is impossible for structures with the inversion symmetry.
 
If the structure has a reflection symmetry in the $y$ direction, i.e. 
\begin{equation*}
\label{eq:eps_yeven} \epsilon(x,y) = \epsilon(x,-y) \quad \mbox{for all} \quad (x,y),
\end{equation*}
then the mapping $y \to -y$ changes $\beta$ to $-\beta$. Thus
\begin{equation*}
\label{eq:Smatrix_yeven} S(\omega, -\beta) = S(\omega, \beta).
\end{equation*}
From Eq.~(\ref{eq:Smatrix_reciprocal}), we have 
\begin{equation}
\label{eq:S_12}   S^{\small \textsf{T}}(\omega, \beta) = S(\omega, \beta).
 \end{equation}
This imples that $t_R = t_L$ for all $\omega$ and  $\beta$.

If the structure has a reflection symmetry in the $x$ direction, i.e. 
\begin{equation*}
\label{eq:eps_xeven} \epsilon(x,y) = \epsilon(-x,y)  \quad \mbox{for all} \quad (x,y),
\end{equation*}
then the mapping $x\to -x$  swaps $a_r$ with $a_l$ and $b_{l0}$ with $b_{r0}$. Thus
\begin{equation}
\label{eq:Smatrix_xeven} P S(\omega, \beta) P^{\small \textsf{T}} = S(\omega, \beta).
\end{equation}
It implies that $r_L = r_R$ and $t_L = t_R$ for all $\omega$ and  $\beta$. Clearly, unidirectional reflectionless is again impossible in this case. Notice that the symmetry in the $x$ direction gives more constrains than each of the other two symmetries studied above.

When $\epsilon$ and $\omega$ are real, the power (per period) carried by the incident wave must equal to the power radiated out by the out-going waves. This leads to the condition that  $S$ must be unitary. A more general result for real $\epsilon$ and complex $\omega$ is
\begin{equation}
\label{eq:Smatrix_Unitarity}  S^*(\bar{\omega}, \beta)  S(\omega, \beta) = I,
\end{equation}
where the superscript \enquote{$^*$} denotes conjugate transpose, $\bar{\omega}$ is the complex conjugate of $\omega$, and $I$ is the identity matrix.  A proof of Eq.~(\ref{eq:Smatrix_Unitarity}) is given in Appendix B. For a real $\omega$, the unitarity of $S$ gives 
\begin{eqnarray*}
|r_L|^2 + |t_L|^2   =  1 \quad \mbox{and} \quad \bar{r}_{L} t_{R} + \bar{t}_{L} r_{R} =  0.
\end{eqnarray*}
Thus, if $r_{L} = 0$, then $|t_L| = 1$ and $r_R = 0$.   Similarly, if $r_{R} = 0$, we must have $r_{L} = 0$. Therefore, unidirectional reflectionless transmission is impossible for lossless dielectric  structures.

If $\epsilon$ is real and symmetric in the $x$ direction, it is known that there could be real frequencies with zero reflections and zero transmissions. Popov {\it et al.} \cite{popov86} first studied a variant of this problem based on the scattering matrix. Shipman and Tu \cite{shipman12} analyzed this problem assuming the structure supports a bound state in the continuum for a nearby $\beta$ and a nearby $\omega$. If one assumes that $\omega^{(r)}$ is a simple (i.e. nondegenerate) zero of $r_L$ (as an analytic function of $\omega$), and it is the only zero in a  domain $\mathcal{W}$ (of the complex $\omega$ plane) containing both $\omega^{(r)}$ and it complex conjugate, then Eq.~(\ref{eq:Smatrix_Unitarity}) allows us to show that $\bar{\omega}^{(r)}$ is a zero of $r_R$. Since $r_L = r_R$ when $\epsilon$ is symmetric in $x$, and $\mathcal{W}$ contains only one zero of $r_L$, we conclude that $\omega^{(r)}$ must be real.

\section{$\mathcal{PT}$-symmetric structures}
\label{sec:PTsym}

We are interested in a class of $\mathcal{PT}$-symmetric structures for which the dielectric function $\epsilon$ is complex and satisfies 
\begin{equation}
\label{eq:PTsymm} \epsilon(x,y) = \bar{\epsilon}(-x,y).
\end{equation}
In other words, the real part of $\epsilon$ is symmetric  and the imaginary part is anti-symmetric in the $x$ direction. In Appendix C, we show that the scattering matrix satisfies
\begin{equation}
\label{eq:Smatrix_PT} P S^*(\bar{\omega}, \beta)  P S(\omega, \beta) = I,
\end{equation}
where $P$ is the $2 \times 2$ matrix given in Sec. \ref{sec:Smatrix} and $I$ is the identity matrix. A more general result (without the parameter $\beta$) on the scattering matrix of $\mathcal{PT}$-symmetric structures is given in \cite{ge12}.

For a real $\omega$, Eq.~(\ref{eq:Smatrix_PT}) gives
\begin{eqnarray}
\label{eq:generalized_unitarity1} && r_L \bar{r}_R  + \bar{t}_R t_L = 1, \\
\label{eq:generalized_unitarity2} && t_R \bar{r}_R + \bar{t}_R r_R = 0, \\
\label{eq:generalized_unitarity3} &&  t_L \bar{r}_L + \bar{t}_L r_L = 0. 
\end{eqnarray}
Let $\phi_1 $ and $\phi_2$ be the  phases of $t_R$ and $t_L$, respectively.  Equation (\ref{eq:generalized_unitarity2})  implies that the  phase of $r_R$ is either $\phi_1 + \pi/2$ or $\phi_1 - \pi/2$. Equation (\ref{eq:generalized_unitarity3}) implies that the  phase of $r_L$ is either $\phi_2 + \pi/2$ or $\phi_2 - \pi/2$.  Thus the phase of $r_R \bar{r}_L$ is $\phi_1 - \phi_2$, or $\phi_1 - \phi_2 \pm \pi$. For all cases, Eq.~(\ref{eq:generalized_unitarity1})  leads to $\phi_1 = \phi_2$.  Therefore $t_R \bar{t}_L$ and $r_L \bar{r}_R$ are real. It is can be verify that $\lambda_1 \lambda_2 = \mbox{det}(S) = - e^{2 i \phi_1}$, where $\lambda_1$ and $\lambda_2$ are the eigenvalues of $S$.

If $t_R \bar{t}_L < 1$,   $r_L \bar{r}_R$ is positive, thus the phases of $r_L$ and $r_R$ are identical, and   Eq.~(\ref{eq:generalized_unitarity1}) leads to
$$ \sqrt{R_L R_R} = 1 - \sqrt{T_L T_R}. $$ 
The above is the generalized energy conservation law \cite{ge12}. If $t_R \bar{t}_L > 1$,    $r_L \bar{r}_R$ is negative, there is a $\pi$ difference between the phases of $r_L$ and $r_R$, and 
Eq.~(\ref{eq:generalized_unitarity1}) leads to
$$ \sqrt{R_L R_R} =  \sqrt{T_L T_R} - 1.$$ 
If $t_R \bar{t}_L = 1$, then at least one of $r_L$ and $r_R$ is zero. If only one of them is zero, we have unidirectional reflectionless transmission. 
For $\beta = 0$ or $\beta \neq 0$ but the structure has an additional reflection symmetry in the $y$ direction, then it is easy to show that $t_L = t_R$. This special case has been extensively studied before \cite{ge12}.

We are interested in $\mathcal{PT}$-symmetric structures that are perturbations of a lossless dielectric structure (with a real dielectric function). If the unperturbed structure has a simple real zero-reflection frequency $\omega_L$  which is the only zero of $r_L$ contained in a domain $\mathcal{W}$ ($\mathcal{W}$ can be chosen as a small disk of the complex $\omega$ plane and centered at $\omega_L$), we expect $r_L$ of the perturbed $\mathcal{PT}$-symmetric structure still has only one simple zero $\tilde{\omega}_L$ in $\mathcal{W}$. We show that $\tilde{\omega}_L$ must still be real. Equation (\ref{eq:Smatrix_PT})  can be written down explicitly as
\begin{eqnarray*}
t_R(\omega, \beta)  \bar{t}_{L}(\bar{\omega}, \beta) +     r_{R}(\omega, \beta)  \bar{r}_{L}(\bar{\omega}, \beta) & =&  1, \\
r_{L}(\omega, \beta) \bar{r}_{L}(\bar{\omega}, \beta)  + t_{L}(\omega, \beta) \bar{t}_{L}(\bar{\omega}, \beta)  & =&  1, \\
t_{R}(\omega, \beta)  \bar{r}_{R}(\bar{\omega}, \beta) + r_{R}(\omega, \beta) \bar{t}_{R}(\bar{\omega}, \beta)  & =&  0, \\
r_{L}(\omega, \beta)  \bar{t}_{L}(\bar{\omega}, \beta) + t_{L}(\omega, \beta)  \bar{r}_{L}(\bar{\omega}, \beta)  & =&  0. 
\end{eqnarray*}
Since $r_L = 0$ at $\omega = \tilde{\omega}_L$,  the last equation  above gives either $r_{L} \left( \bar{\tilde{\omega}}_L, \beta \right) = 0$ or $t_{L} \left( \tilde{\omega}_L , \beta \right) = 0$. But the second equation above indicates that $t_{L} \left( \tilde{\omega}_L, \beta \right) $ can not be zero,  thus $r_{L} \left( \bar{\tilde{\omega}}_L, \beta \right) = 0$. Since the domain $\mathcal{W}$ only contains only one zero of $r_L$, we must have  $\tilde{\omega}_L = \bar{\tilde{\omega}}_{L}$, i.e. $\tilde{\omega}_{L}$ is real.  Similarly,  if we make proper assumptions about a right real zero-reflection frequency, $\omega_R$, of the unperturbed structure, we can show that there must be a real zero, $\tilde{\omega}_R$, of the right reflection coefficient $r_R$ for the perturbed $\mathcal{PT}$-symmetric structure.

\section{Perturbation analysis}
\label{sec:perturbation}
The previous section establishes  the continual existence of real zero-reflection frequencies for both the left and right incident waves under $\mathcal{PT}$-symmetric perturbations. In this section,  we use a perturbation method to show that these two frequencies are different in general, and thus unidirectional reflectionless transmission indeed occur. First, we consider a lossless dielectric structure with a real dielectric function $\epsilon$ that is also symmetric in the $x$ direction, and assume that there is a real wavenumber $\beta$ and  real zero-reflection frequency $\omega_L = \omega_R$ for both left and right incident waves.   Let $u_L$ and $u_R$  be the corresponding diffraction solutions for left and right incident plane waves with unit amplitude, respectively. The symmetry of the structure in $x$ implies that $u_R(x,y) = u_L(-x,y)$.  At $x = \pm D$, these two solutions can be written down as
\begin{eqnarray}
\label{eq:uL_xpmD_fourier} &&  u_R (D,y) = u_L(- D,y) =  e^{i \beta y} + \sum\limits_{j=-\infty}^{\infty} b^{-}_j e^{i \beta_j y}, \\ 
&& u_R(-D,y) = u_L( D,y) =   \sum\limits_{j=-\infty}^{\infty} b^{+}_j e^{i \beta_j y},
\end{eqnarray} 
where $\{ b^{+}_j \}$ are the Fourier coefficients of $u_L( D, y)$, and $\{ b^{-}_j \}$ are the Fourier coefficients of $u_L( - D, y) - e^{i \beta y}$. Since the reflection is zero and  energy is conserved, we have
$$b^-_0 = 0, \quad  \left| b^+_0 \right| = 1.$$

Next, we consider a perturbed $\mathcal{PT}$-symmetric  structure with a dielectric function
\begin{equation}
\label{eq:eps_perturbation}
 \tilde{\epsilon} = \epsilon +   \delta  F(x,y), 
 \end{equation}
where $\delta$ is  a small real number,  $F$ is a complex $O(1)$ function  satisfying 
$$ F(x,y) = \bar{F}(-x,y),$$
and $F(x,y) = 0$ for $|x| > D$. According to Sec.~{\ref{sec:PTsym}}, there must be  real frequencies  $\tilde{\omega}_L$ and $\tilde{\omega}_{R}$ such that $r_L = 0$ and $r_R = 0$, respectively, for the perturbed structure $\tilde{\epsilon}$  and the fixed $\beta$.
 
Let $\tilde{k}_L = \tilde{\omega}_L/ c$ and $\tilde{k}_R = \tilde{\omega}_R / c$,  where $c$ is the speed of light in vacuum. We expand  $\tilde{k}_L$ and $\tilde{k}_R$  in power series of $\delta$:
\begin{eqnarray}
\label{eq:exp_freq1} \tilde{k}_L &= & k_0 + k^{(L)}_{1} \delta + k^{(L)}_{2} \delta^2 + \ldots, \\
\label{eq:exp_freq2} \tilde{k}_R & = & k_0 + k^{(R)}_{1} \delta + k^{(R)}_{2} \delta^2 + \ldots,
\end{eqnarray}
where $k_0 = \omega_L / c = \omega_R/c$. In Appendix D, we show that the  coefficients $k_1^{(L)}$ and $k_1^{(R)}$ are given by
\begin{widetext}
\begin{equation}
\label{eq:k1_left} k^{(L)}_1 = \dfrac{ -  k_0 \int_{\Omega_D} F u_L \bar{u}_R   dxdy}{   \dfrac{i L}{\alpha} (b^+_0 - \bar{b}^+_0) + i L \sum\limits_{\substack{j=-\infty \\ j \neq 0}}^{\infty} \dfrac{\bar{b}^-_j b^+_j + \bar{b}^-_j b^+_j}{\alpha_j} + 2  \int_{\Omega_D} \epsilon u_L \bar{u}_R   dxdy },
\end{equation}
and
\begin{equation}
\label{eq:k1_right} k^{(R)}_1 = \dfrac{ -  k_0 \int_{\Omega_D} F \bar{u}_L u_R  dxdy}{   \dfrac{i L}{\alpha} (b^+_0 - \bar{b}^+_0) + i L \sum\limits_{\substack{j=-\infty \\ j \neq 0}}^{\infty} \dfrac{\bar{b}^-_j b^+_j + \bar{b}^-_j b^+_j}{\alpha_j} + 2 \int_{\Omega_D} \epsilon \bar{u}_L u_R  dxdy  },
\end{equation}
\end{widetext}
where $\Omega_D $ is the rectangle given by $|x| < D$ and $|y| < L/2$. The coefficients $\{ \alpha_j \}$ are defined in Sec.~\ref{sec:Smatrix}, $\omega_L$ and $\beta$ are assumed such that  $\alpha = \alpha_0$ is real, and all other $\alpha_j$ for $j \neq 0$ are pure imaginary. It is easy to verify that  $\bar{u}_R u_L$ is a  $\mathcal{PT}$-symmetric function satisfying the same Eq.~(\ref{eq:PTsymm}). Using this result and the symmetry of $\epsilon$ and $F$,  it is straightforward to show  that $k^{(L)}_1$ and $k^{(R)}_1$ are real, and the denominators of Eqs.~(\ref{eq:k1_left}) and (\ref{eq:k1_right}) are identical. Therefore,  if  $F(x,y)$ satisfies 
\begin{equation}
\label{eq:cond_F}   \int_{\Omega_D} \left[ F(x,y) - F(-x,y) \right] u_L \bar{u}_R   dxdy \neq  0, 
\end{equation}
then $k^{(L)}_1 \neq k^{(R)}_1$.  This implies that as far as $F$ satisfies Eq.~(\ref{eq:cond_F}), $\tilde{\omega}_L \neq \tilde{\omega}_R$ for arbitrarily small $\delta$,  and unidirectional reflectionless transmission occurs at frequencies $\tilde{\omega}_L $ and $\tilde{\omega}_R$ for left and right incident waves, respectively.

\section{Numerical examples}
\label{sec:Numerical}
\begin{figure}[http]
\centering
\includegraphics[scale=0.65]{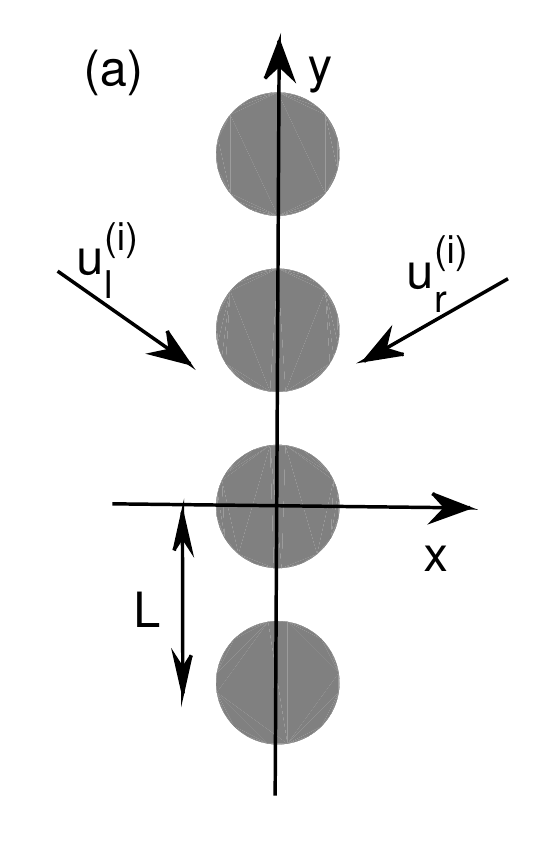}
\includegraphics[scale=0.65]{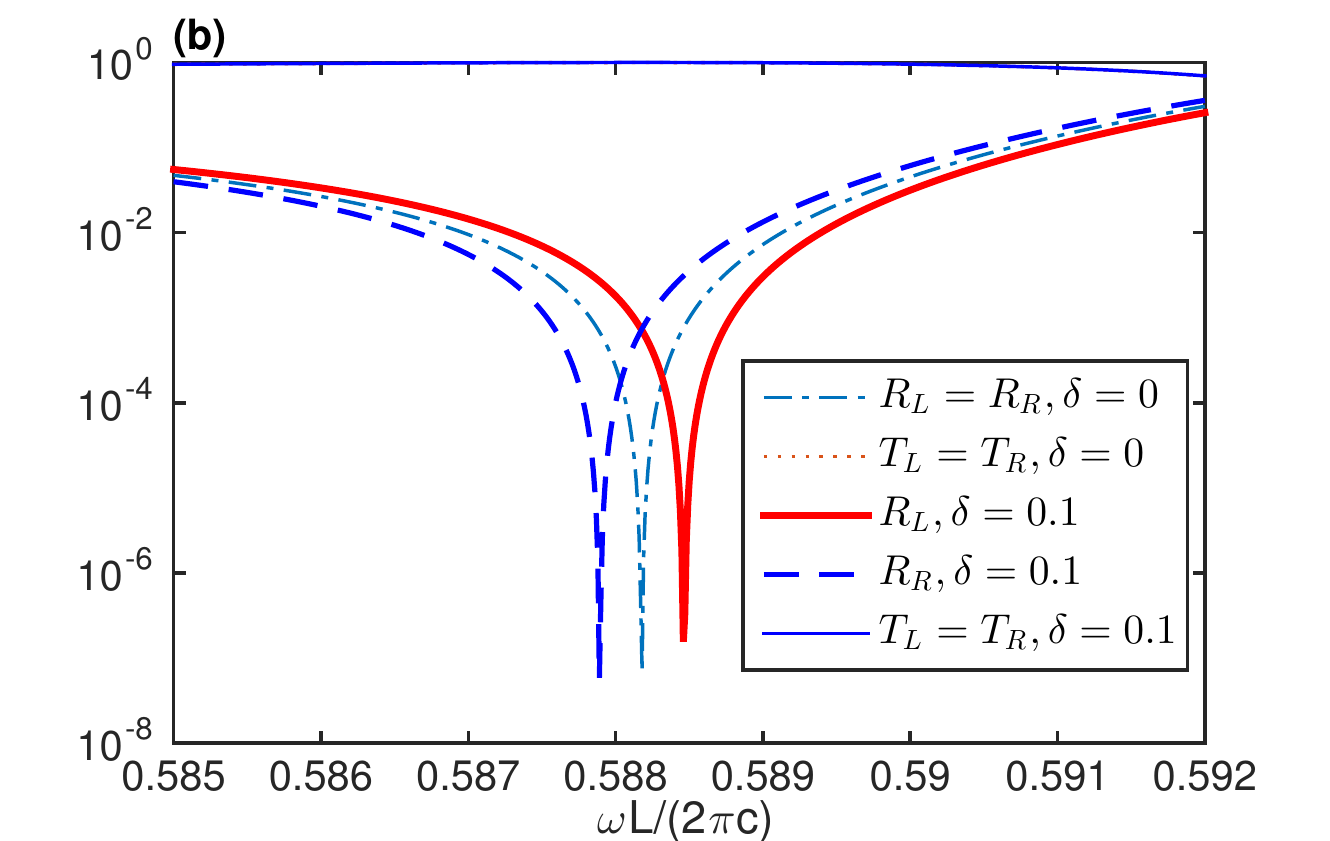}
\caption{(a) A periodic array of identical circular cylinders with incident waves from left and right. (b) Example 1: reflection and transmission spectra  for normal incident waves.  }
\label{fig:fig2}
\end{figure}

In this section, we present numerical results to illustrate the unidirectional reflectionless transmission phenomenon, validate the perturbation results of Sec.~\ref{sec:perturbation}, and show examples of unidirectional invisibility.  We consider a periodic array of identical circular cylinders with period $L$ in the $y$ direction and surrounded by air as shown in Fig.~{\ref{fig:fig2}}(a). The coordinates are chosen so that the centers of the cylinders are on the $y$ axis and the center of one cylinder is at the origin. 
The first example is for cylinders with a $y$-independent dielectric function given by
\begin{equation}
\label{eq:dielectric_cylinder} \epsilon(x,y) =  \epsilon_1 + i \delta \sin\left(\dfrac{\pi x}{2 a} \right), 
\end{equation}
 where $a = 0.3 L$ is the radius of the cylinders,  $\epsilon_1 = 10$, and  $ \delta$ is a real parameter.    If $\delta \neq 0$, the structure is $\mathcal{PT}$-symmetric with respect to the reflection  in the $x$ direction. The diffraction problem for a given incident wave can be solved by many different numerical methods. We use a mixed Fourier-Chebyshev pseudospectral method \cite{tref00} to discretize the Helmholtz equation inside the disk of a radius $a$ (corresponding to the cross section of the central cylinder), and use cylindrical and plane wave expansions outside the cylinders \cite{huang06}. In Fig.~{\ref{fig:fig2}}(b), we show the reflection and transmission spectra in logarithmic scale for left and right  normal incident waves.  For $\delta = 0$, $\epsilon$ is real and  symmetric in both $x$ and $y$ directions, thus the reflection and transmission spectra are identical for left and right  incident waves, i.e. $R_L = R_R$ and $T_L = T_R$. In Fig.~{\ref{fig:fig2}}(b), a dip can be observed in the reflection spectrum for $\delta  =0$. Presumably, the reflection coefficient is exactly zero at normalized frequency $\omega L /(2\pi c) \approx 0.5882$. To use the perturbation results of Sec.~\ref{sec:perturbation}, we assume $D = L/2$, then $\Omega_D$ is a square of side length $L$ centered at the origin. Using the numerical solutions to evaluate Eqs.~(\ref{eq:k1_left}) and (\ref{eq:k1_right}), we obtain $k^{(L)}_1 \approx 0.018 L^{-1}$ and $k^{(R)}_1 \approx  -0.018 L^{-1}$. This implies that when $\delta$ is increased from zero, the zero-reflection frequencies for left and right  incident waves will increase and  decrease, respectively. The numerical results for $\delta = 0.1$ are also shown in Fig.~{\ref{fig:fig2}}, and they indicate  that the normalized zero-reflection frequency $\omega L/(2\pi c)$ is increased  to $ 0.5885$ for left incident waves, and is  decreased to  $0.5879$ for right incident waves. According to our perturbation theory, the normalized  zero-reflection frequency  to the left incident waves is approximately $0.5882 + \delta k^{(L)}_1 L /(2\pi)$. Keeping four significant digits, this value is also $0.5885$.  Therefore, the numerical and perturbation results agree very well with each other.    At the zero-reflection frequency for left (or right) incident waves, the reflection coefficient for right (or left) incident waves is non-zero,  therefore, we have unidirectional reflectionless transmissions. Since the structure is symmetric in $y$, the transmission coefficients for left and right incident waves are identical. From Eq.~(\ref{eq:generalized_unitarity1}), it is clear that the transmittance is exactly $1$ at the zero-reflection frequencies. 

\begin{figure}[http]
\centering
\includegraphics[scale=0.55]{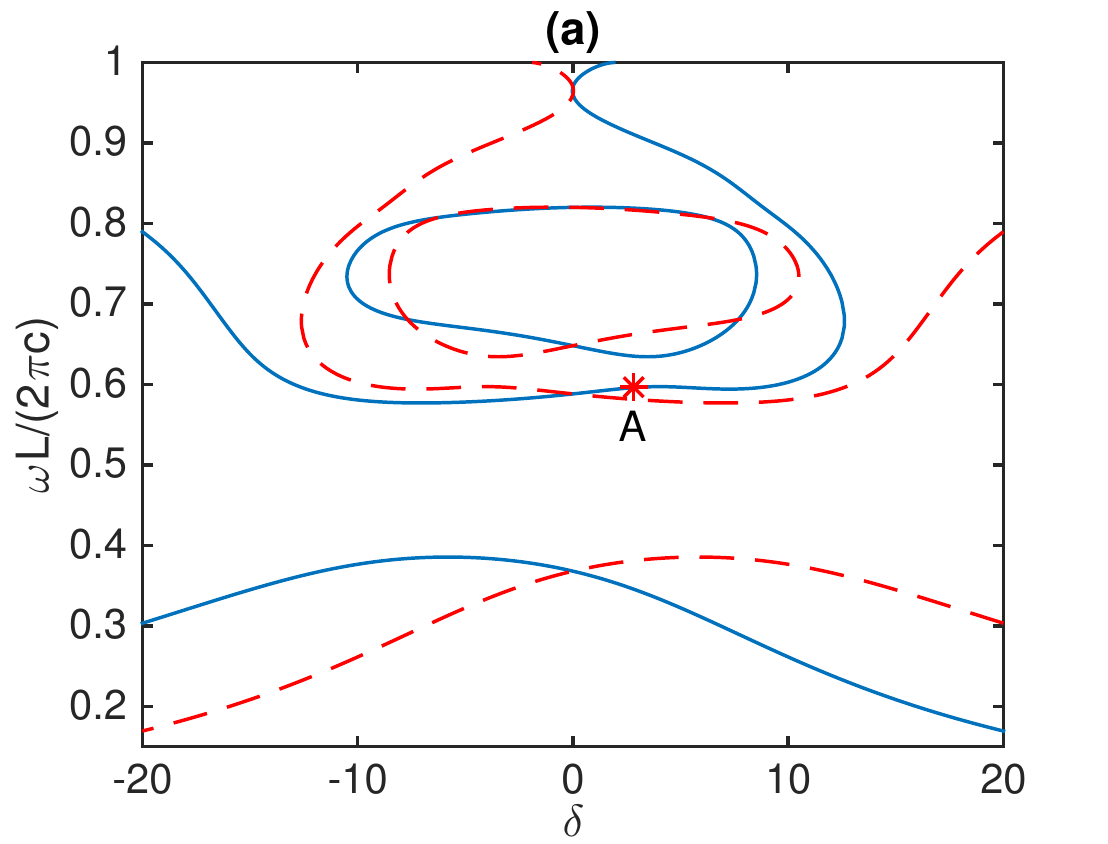}
\includegraphics[scale=0.55]{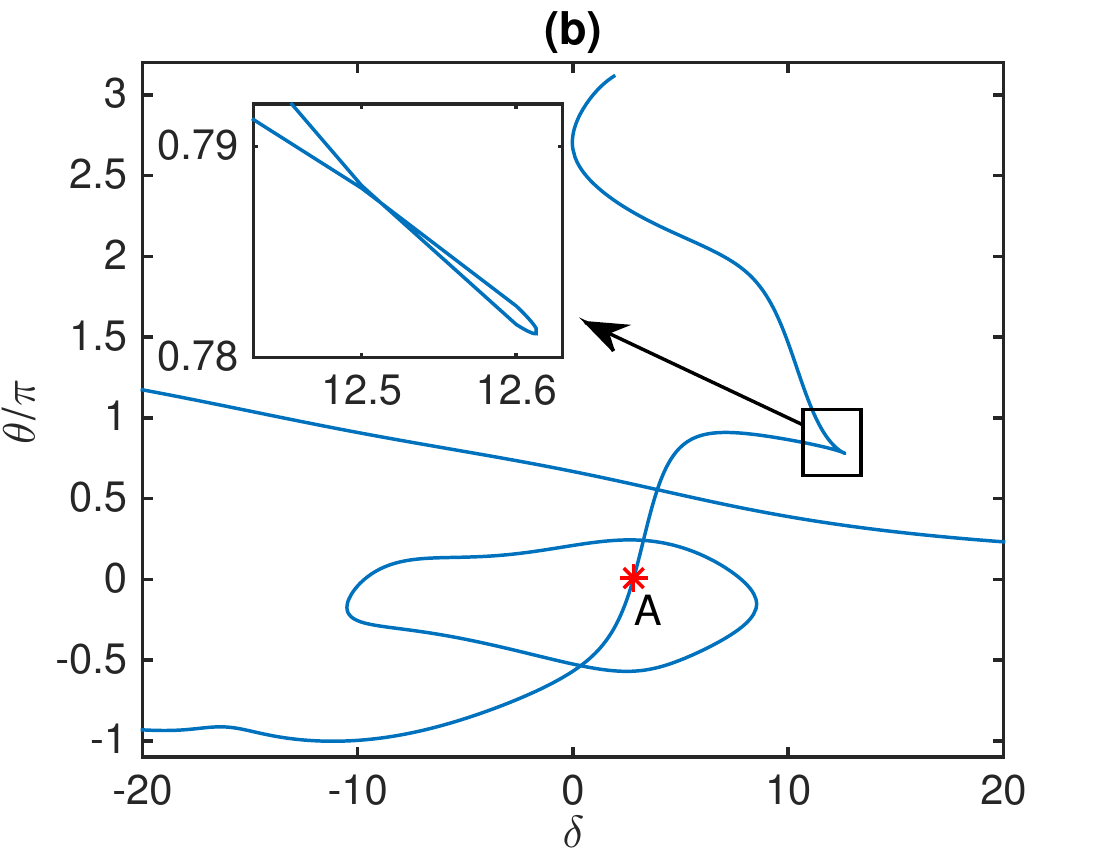}
\caption{(a) Zero-reflection frequencies for different $\delta$ in the first example.  Blue solid line: left incident wave, red dashed line: right incident waves. (b) Relative phases $\theta$ of the transmitted waves at   zero-reflection frequencies for left  incident waves.  }
\label{fig:fig3}
\end{figure}

\begin{figure}[http]
\centering
\includegraphics[scale=0.5]{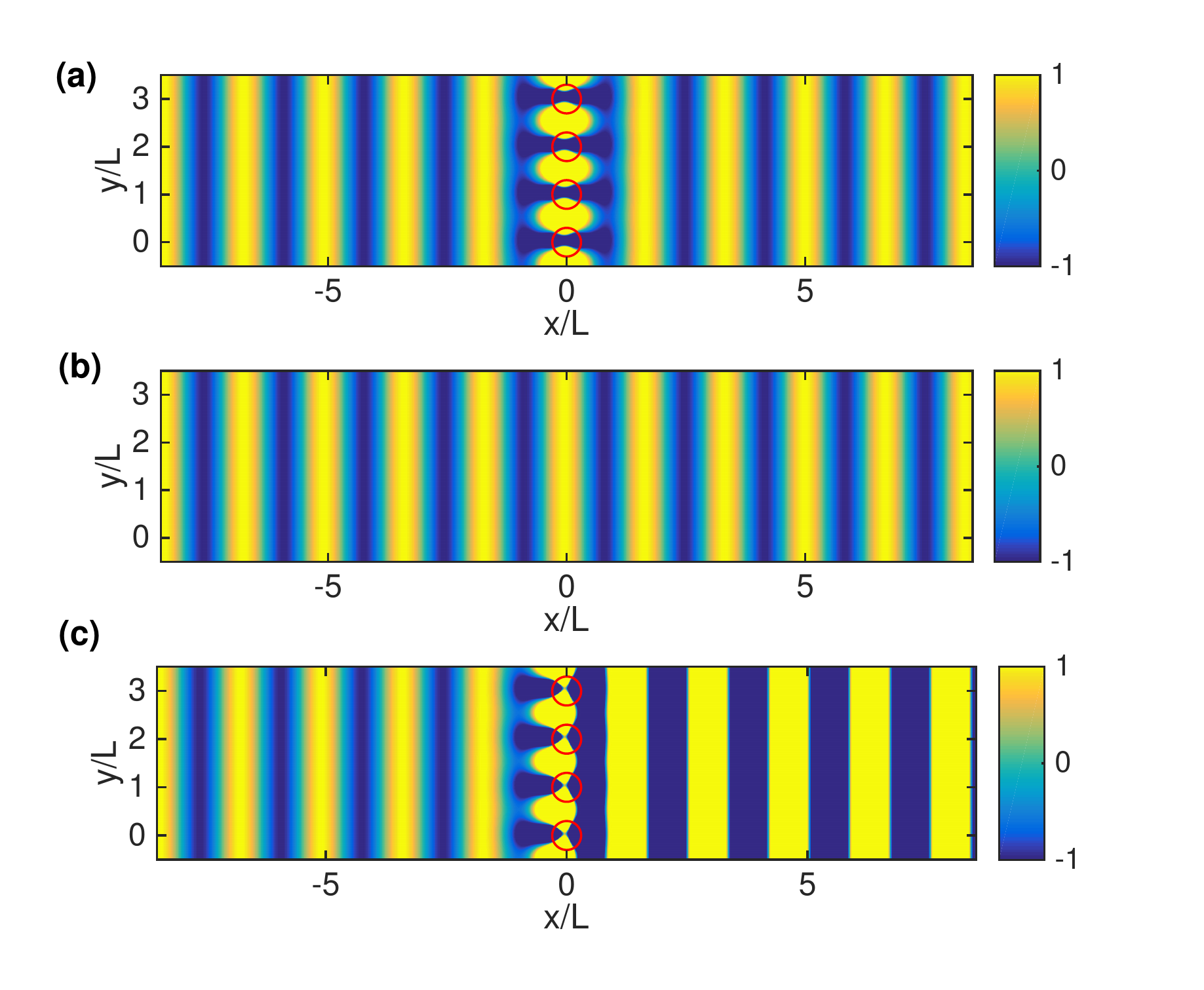}
\caption{Real part of  the diffraction solutions, i.e. Re$(u)$, corresponding to point $A$ in Fig.{\ref{fig:fig3}}.   (a) Left  incident wave with $a_l=e^{-i \alpha D }$ and $a_r=0$. (b) Plane wave $e^{i k_0 x}$ propagating in air.  (c) Right incident wave with $a_l=0$ and $a_r= e^{i \alpha D}$. Red circles denote the cylinders. The wave fields are capped from $-1$ to $1$. }
\label{fig:fig4}
\end{figure}

In Fig.~{\ref{fig:fig3}}(a) we show  zero-reflection frequencies for different $\delta$. The solid blue curve and the dashed red curve correspond to left and right incident waves, respectively.  For $\delta=0$, the  zero-reflection frequencies for left and right  incident waves are identical. For $\delta \neq 0$, they are different in general, thus unidirectional reflectionless transmission can occur. Overall, the curves for left and right  zero-reflection frequencies are mirror images of each other. At a zero-reflection frequency for a left incident wave, the transmission coefficient $r_L$ has unit magnitude, and we can define a phase $\theta$ relative to the incident wave (extended to the whole space) by
$$ e^{i \theta} = t_L e^{- 2 i \alpha D}. $$
 In Fig.~\ref{fig:fig3}(b), we show the relative phase $\theta$ for all  zero-reflection frequencies on the blue solid curves in Fig.~{\ref{fig:fig3}}(a).
 Point $A$ in Figs.~{\ref{fig:fig3}}(a) and (b) is a point with a zero relative phase. It is obtained at $\omega L /(2\pi c) = 0.5959$ for $ \delta = 2.8$.  In Figs.~{\ref{fig:fig4}}(a) and (c), we show  the diffraction solutions for left and right  normal incident waves for point $A$.  For a left incident wave (i.e. $a_l=e^{- i \alpha D}, a_r=0$), the total  wave  is identical to the  plane wave $e^{i k_0 x}$ (shown in Fig.~{\ref{fig:fig4}}(b)) away from the cylinders.  This implies that the cylinders are invisible to left incident waves.   For a right incident wave (i.e. $a_l=0, a_r= e^{i \alpha D}$),  the transmitted wave also has unit amplitude and zero relative phase, the reflected wave has a magnitude about $6.93$.  Although the $\mathcal{PT}$-symmetric structure has a balanced gain and loss profile, energy does not need to be conserved. For this case, a strong reflected wave is produced thanks to the gain medium. 

\begin{figure}[http]
\centering
\includegraphics[scale=0.55]{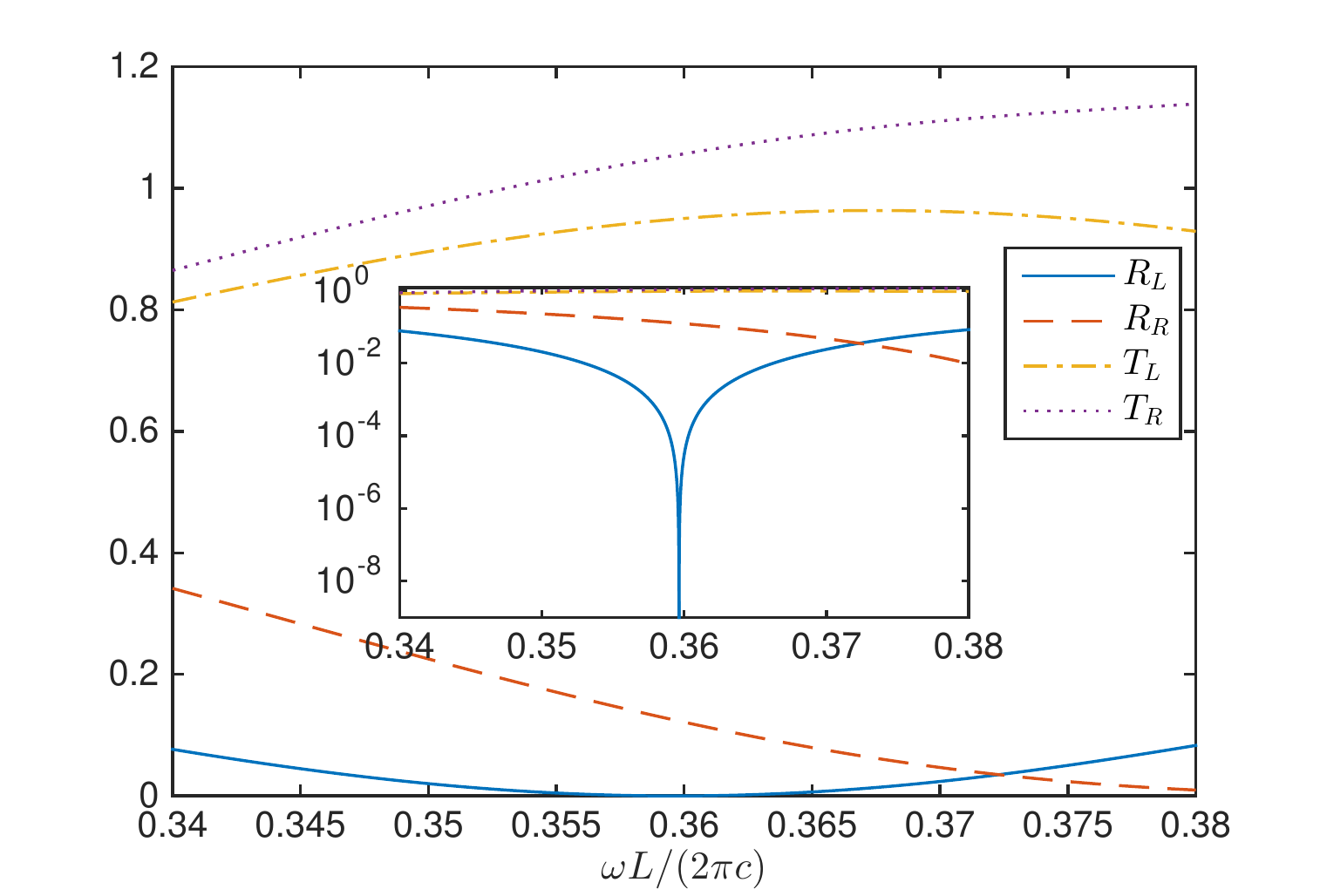}
\caption{Example 2: reflection and transmission spectra for $\beta L /(2\pi) = 0.2$.   Insert is the logarithmic scale plot.}
\label{fig:fig_example2}
\end{figure}

\begin{figure}[http]
\centering
\includegraphics[scale=0.62]{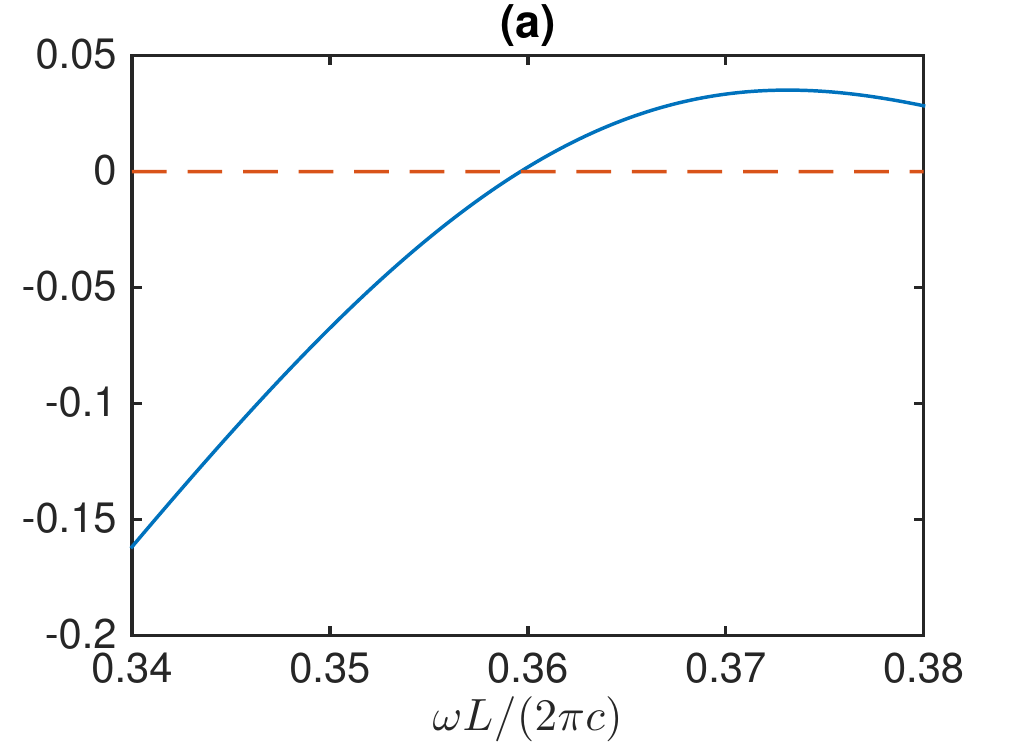}
\includegraphics[scale=0.62]{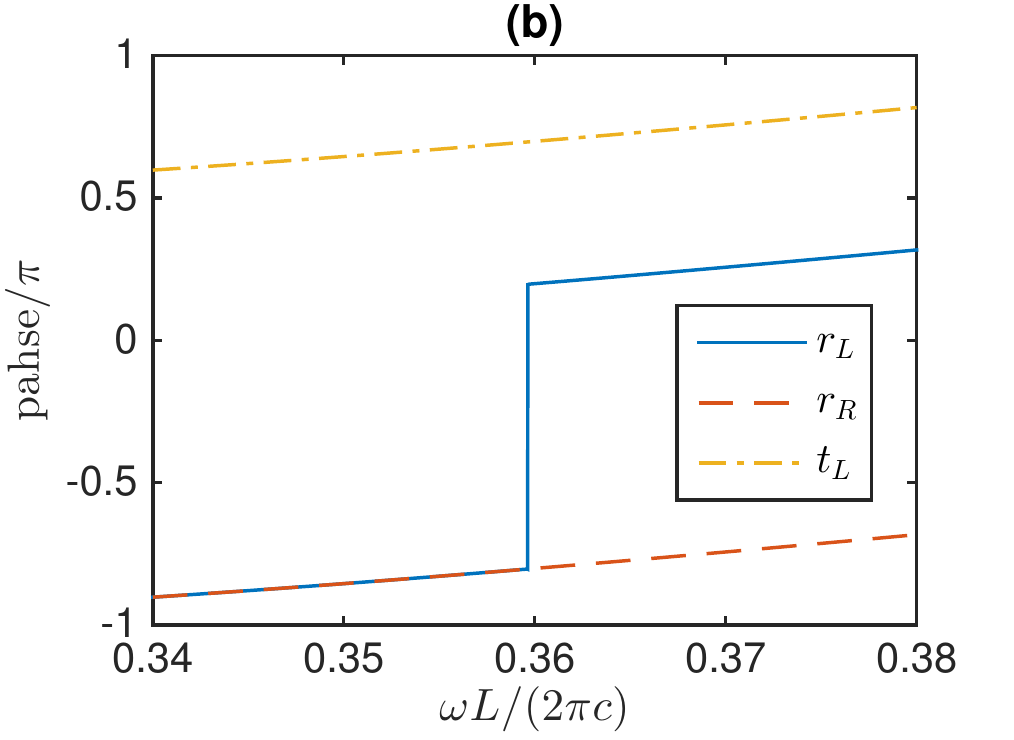}
\caption{ (a) Real (solid line) and imaginary (dashed line) parts of $t_R \bar{t_R} - 1$. (b) Normalized phases of $r_L$ (solid line), $r_R$ (dashed line) and $t_L$ (dashdot line).}
\label{fig:fig_example2_2}
\end{figure}

In order to study $\mathcal{PT}$-symmetric structures without the reflection symmetry in $y$, we consider another example. The  structure is again a periodic array of identical circular cylinders with their centers on the $y$ axis,   but the  dielectric function of the cylinder centered at the origin  is  given by
\begin{equation}
 \epsilon(x,y) =  \epsilon_1 + 2  \sin\left(\dfrac{\pi y}{2 a} \right) + 2 i  \sin\left(\dfrac{\pi x}{2 a} \right),
\end{equation}
where $\epsilon_1$ and $a$ are the same as the first example.   The structure is $\mathcal{PT}$-symmetric with respect to a reflection in the $x$ direction, but it is not  symmetric in $y$. For $\beta \neq 0$, the transmission coefficients for left and right incident waves are different in general. In Fig.~{\ref{fig:fig_example2}}, we show the reflection and transmission spectra for left and right incident waves with $\beta L/(2 \pi)= 0.2$. At $\omega L/(2\pi c) = 0.3569$, the reflection is zero  for left incident waves and nonzero for right incident waves, thus unidirectional reflectionless transmission occurs. The corresponding transmittance are $T_L = 0.9488$ and $T_R = 1.0539$. From Sec.~\ref{sec:PTsym}, we know that $t_R \bar{t}_L$ is aways real. To verify this,   we show the value of $t_R \bar{t}_L - 1$ as a function of the frequency in Fig.~{\ref{fig:fig_example2_2}}(a). Clearly,  $t_R \bar{t}_L = 1 $ at zero-reflection frequency $\omega L/(2\pi c) = 0.3569$. In Fig.~{\ref{fig:fig_example2_2}}(b), we show the phases of $r_L$ and $r_R$. Notice that the phase of $r_L$ has a jump discontinuity of $\pi$ at the the zero-reflection frequency.

\section{Conclusions}

In this paper, we studied   2D $\mathcal{PT}$-symmetric periodic structures sandwiched between two homogeneous media.  Using a scattering matrix  formalism  we showed that the real zero-reflection frequencies are robust under $\mathcal{PT}$-symmetric perturbations. A simple condition on the perturbed dielectric function was derived by a perturbation method to guarantee that the real zero-reflection frequencies for left and right incident waves are different. Therefore, as far as the original unperturbed dielectric structure is symmetric and has a real non-degenerate zero-reflection frequency, unidirectional reflectionless transmission is certain to occur for almost any $\mathcal{PT}$-symmetric perturbations. Numerical examples are presented for periodic arrays of circular cylinders with $\mathcal{PT}$-symmetric dielectric functions. The numerical results confirmed the perturbation theory, and illustrated unidirectional invisibility and other interesting wave phenomena. 

Both the scattering matrix formalism and the perturbation analysis can be easily extended to unperturbed structures that are themselves $\mathcal{PT}$-symmetric. The scattering matrix formalism allows us to conclude that real non-degenerate zero-reflection frequencies are protected by the $\mathcal{PT}$-symmetry, in the sense that these frequencies remain real for any $\mathcal{PT}$-symmetric perturbations. The perturbation theory gives quantitative results on the changes of the real  zero-reflection frequencies caused by perturbations.  It is also straightforward to consider non-$\mathcal{PT}$-symmetric perturbation that could move the real zero-reflection frequencies to the complex plane. Our study enhances the theoretical understanding on the  zero-reflection frequencies, unidirectional reflectionless transmission, and unidirectional invisibility for $\mathcal{PT}$-symmetric structures, and provides a solid foundation for further studies on these wave phenomena and for exploring their potential applications.

\section*{Acknowledgments}
The authors acknowledge support from the Science and Technology Research Program of Chongqing
Municipal Education Commission, China (Grant No. KJ1706155),  and the Research Grants
Council of Hong Kong Special Administrative Region, China (Grant
No. CityU 11304117).

\section*{Appendix}
\subsection*{Appendix A: Reciprocity}
To derive Eq.~(\ref{eq:Smatrix_reciprocal}), we consider the diffraction problem for incident waves $\tilde{a}_{l} e^{i [ \alpha (x +D) - \beta y]}$ and $\tilde{a}_r e^{-i[\alpha (x - D) + \beta y]}$ with frequency $\omega$ and wavenumber $-\beta$.   The diffraction solution $\tilde{u}$ can be written as
\begin{equation*}
\tilde{u}(x,y) =  \tilde{a}_l e^{i [\alpha (x + D) - \beta y]}  +  \sum\limits_{j=-\infty}^{+\infty} \tilde{b}_{lj} e^{- i [ \tilde{\alpha}_j (x + D) - \tilde{\beta}_j y]}, 
\end{equation*}
for $ x < - D$ and
\begin{equation*}
 \tilde{u}(x,y) =       \tilde{a}_r e^{- i [ \alpha (x - D) + \beta y]} + \sum\limits_{j=-\infty}^{+\infty} \tilde{b}_{rj} e^{i [ \tilde{\alpha}_j (x - D) + \tilde{ \beta}_j y]},  
\end{equation*}
for $x > D$,
where 
$$ \tilde{\beta}_j = - \beta + 2 j \pi / L, \quad \tilde{\alpha}_j = \sqrt{k_0^2 - \tilde{\beta}_j^2},  $$
for $ j=0, \pm1, \pm 2, \ldots$. Notice that 
$$ \tilde{\beta}_j = - \beta_{-j}, \quad \tilde{\alpha}_j = \alpha_{-j}, \quad  \mbox{for} \quad j=0, \pm1, \pm 2, \ldots. $$
The coefficients $\tilde{b}_{l0}$ and $\tilde{b}_{r0}$  are related to the incident coefficients $\tilde{a}_l$ and $\tilde{a}_r$  by scattering matrix $S(\omega, - \beta)$ as
$$
 \left[ \begin{matrix} \tilde{b}_{l0} \\ \tilde{b}_{r0}    \end{matrix} \right] = S(\omega, -\beta) \left[ \begin{matrix} \tilde{a}_{l} \\ \tilde{a}_{r}    \end{matrix} \right].$$

From the governing equations of $u$ and $\tilde{u}$, we have
$$ 0 = \tilde{u} (\Delta u + k^2_0 \epsilon u) - u (\Delta \tilde{u} + k_0^2 \epsilon \tilde{u}) = \nabla \cdot (\tilde{u} \nabla u) - \nabla \cdot ({u} \nabla \tilde{u}),$$
where $\Delta = \partial^2_x + \partial^2_y$ is the Laplace operator. Integrating the above equation on domain $\Omega_D$, we have
$$ \int_{\partial \Omega_D}  \left( \tilde{u} \dfrac{\partial u}{\partial \nu} - {u} \dfrac{\partial \tilde{u}}{\partial \nu}    \right) ds= 0.$$
The line integrals on the two edges of $\Omega_D$ at $y = \pm L/2$ cancel out. From the expressions of $u$ (i.e. Eqs.~(\ref{eq:scat_solution_xlessD}) and (\ref{eq:scat_solution_xlargeD})) and $\tilde{u}$ at $x = \pm D$, and the relations of $\beta_j$ and $\tilde{\beta}_j$, we obtain
$$ \tilde{a}_l b_{l0} + \tilde{a}_r b_{r0} = a_l \tilde{b}_{l0} + a_r \tilde{b}_{r0}. $$
Therefore, 
$$  \left[ a_l,  a_r  \right] \left[ S^{\small \textsf{T}}(\omega, \beta) - S(\omega, -\beta) \right] \left[ \begin{matrix} \tilde{a}_{l} \\ \tilde{a}_r  \end{matrix} \right]  = 0$$
for any complex $a_l, a_r, \tilde{a}_l$ and $\tilde{a}_r$. This leads to Eq.~(\ref{eq:Smatrix_reciprocal}).

\subsection*{Appendix B: Unitarity  for lossless dielectric structures}
\label{App:unitarity}
Assume  $\epsilon$ is real and $\omega$ is complex with a small (positive and negative) imaginary part. We consider the diffraction problem for two incident waves $\hat{a}_l e^{i [\alpha(\bar{\omega}) (x + D)+ \beta y]}$ and $\hat{a}_r e^{-i [\alpha(\bar{\omega}) (x - D) - \beta y]}$ with frequency $\bar{\omega}$ and wavenumber $\beta$, where $\alpha(\bar{\omega})  = \sqrt{\bar{k}_0^2 - \beta^2}$ and $k_0 = \omega / c$. Here the square root is defined with a branch cut on the negative imaginary axis as shown in Sec.~\ref{sec:Smatrix}.

The  diffraction problem is governed by the Helmholtz equation 
\begin{eqnarray*}
\Delta w + \bar{k}_0^2 \epsilon w = 0.
\end{eqnarray*}
 The solution $w$ can be written as
\begin{equation*}
 w(x,y)=  \hat{a}_l e^{i[ \alpha(\bar{\omega}) (x + D) + \beta y]} +   \sum\limits_{j = -\infty}^{+\infty} \hat{b}_{lj} e^{-i [  \alpha_j(\bar{\omega}) (x + D) - {\beta}_j y ]},  
\end{equation*}
for $x < - D$ and
\begin{equation*}
 w(x,y)=   \hat{a}_r e^{- i [ \alpha(\bar{\omega}) (x - D) - \beta y]}  + \sum\limits_{j=-\infty}^{+\infty} \hat{b}_{rj} e^{i  [ \alpha_j(\bar{\omega}) (x - D)  +  { \beta}_j y]},  
\end{equation*}
for $x > D$, 
where
$$ \alpha_j(\bar{\omega}) = \sqrt{\bar{k}_0^2 - \beta_j^2}, \quad \mbox{for} \quad j = 0, \pm 1, \pm 2, \ldots. $$
 The relations between $\alpha_j(\bar{\omega})$ and $\alpha_j(\omega)$ are
$$ \alpha_j(\bar{\omega}) = \left\{  \begin{array}{ll} \bar{\alpha}_j(\omega), & j = 0 \\
                                                                                 - \bar{\alpha}_j(\omega), & j = \pm 1, \pm 2, \ldots.
                        \end{array} \right. $$
 The coefficients $\hat{b}_{l0}$ and $\hat{b}_{r0}$  are related to the incident coefficients $\hat{a}_l$ and $\hat{a}_r$  by scattering matrix  $S(\bar{\omega}, \beta)$ as
\begin{equation}
\label{eq:Smatrix_complexFreq}
 \left[ \begin{matrix} \hat{b}_{l0} \\ \hat{b}_{r0}    \end{matrix} \right] = S(\bar{\omega}, \beta) \left[ \begin{matrix} \hat{a}_{l} \\ \hat{a}_{r}    \end{matrix} \right].
 \end{equation}

 From the governing equations of $u$ and $w$,  we have
$$ 0 = \bar{w}  (\Delta u + k^2_0 \epsilon u) - u (\Delta \bar{w} + k_0^2 \epsilon v) = \nabla \cdot (\bar{w} \nabla u) - \nabla \cdot ({u} \nabla \bar{w}).$$
Integrating the above equation on domain $\Omega_D$, we have
$$ \int_{\partial \Omega_D}  \left( \bar{w} \dfrac{\partial u}{\partial \nu} - {u} \dfrac{\partial \bar{w}}{\partial \nu}    \right) ds= 0.$$
The line integrals on the two edges of $\Omega_D$ at $y = \pm L/2$ cancel out. From the expressions of $u$  and $w$ at $x = \pm D$, and the relations of $\alpha_j$ and $\alpha_j(\bar{\omega})$, we obtain
$$  a_{l}\bar{\hat{a}}_l  + a_{r} \bar{\hat{a}}_r  = b_{l0}\bar{\hat{b}}_{l0} +  b_{r0}\bar{\hat{b}}_{r0} . $$
Therefore, 
$$  \left[ \bar{\hat{a}}_{l} , \bar{\hat{a}}_r  \right] \left[ S^*(\bar{\omega}, \beta)  S(\omega, \beta)   - I \right] \left[ \begin{matrix} a_{l} \\ a_r  \end{matrix} \right]  = 0$$
for any complex $a_l, a_r, \hat{a}_l$ and $\hat{a}_r$. This leads to Eq.~(\ref{eq:Smatrix_Unitarity}).

\subsection*{Appendix C: Scattering matrix for $\mathcal{PT}$-symmetric structures}

Let  $\epsilon$ satisfy the $\mathcal{PT}$-symmetric condition Eq.~(\ref{eq:PTsymm}),  $\omega$ be complex with a small (positive or negative) imaginary part,  $w$ be the diffraction solution  defined in Appendix B and $v = \bar{w}(-x,y)$, then 
 $$ \Delta v + k_0^2 \epsilon v = 0.$$

From the governing equations of $u$ and $v$, we have
$$ 0 = v  (\Delta u + k^2_0 \epsilon u) - u (\Delta v + k_0^2 \epsilon v) = \nabla \cdot (v \nabla u) - \nabla \cdot ({u} \nabla v).$$
Integrating the above equation on domain $\Omega_D$, we have
$$ \int_{\partial \Omega_D}  \left( v \dfrac{\partial u}{\partial \nu} - {u} \dfrac{\partial v}{\partial \nu}    \right) ds= 0.$$
The line integrals on the two edges of $\Omega_D$ at $y = \pm L/2$ cancel out. Using the expressions of $u$ and $v$ (notice that $v = \bar{w}(-x,y)$), we have
$$  a_{l}\bar{\hat{a}}_r  + a_{r} \bar{\hat{a}}_l  = b_{l0}\bar{\hat{b}}_{r0} +  b_{r0}\bar{\hat{b}}_{l0} . $$
From Eqs.~(\ref{eq:Smatrix}) and (\ref{eq:Smatrix_complexFreq}), we obtain
$$  \left[ \bar{\hat{a}}_{r} , \bar{\hat{a}}_l   \right] \left[ P S^*(\bar{\omega}, \beta) P S(\omega, \beta)   - I \right] \left[ \begin{matrix}  a_l \\  a_r \end{matrix} \right]  = 0$$
for any complex $a_l, a_r, \hat{a}_l$ and $\hat{a}_r$. This leads to Eq.~(\ref{eq:Smatrix_PT}). 
 
 \subsection*{Appendix D: Perturbation analysis}
 To carry out the perturbation analysis, we first formulate a boundary value problem for the diffraction problem. Notice that a diffraction solution for  incident waves given in Eqs.~(\ref{eq:incident_left}) and (\ref{eq:incident_right}) has the general expansions given in Eqs.~(\ref{eq:scat_solution_xlessD}) and ~(\ref{eq:scat_solution_xlargeD}) for $|x| > D$. If  we define a linear operator $\mathcal{T}$  such that
\begin{equation}
\label{eq:T}
\mathcal{T} e^{i \beta_j y} = i \alpha_j e^{i \beta_j y}, \quad j=0, \pm 1, \pm2, \ldots, 
\end{equation}
then the diffraction solution $u$ satisfies the following boundary conditions \cite{bao95}
\begin{equation}
\label{eq:BCxmpD}
\left\{ \begin{array}{ll} \dfrac{\partial u}{\partial x} = - \mathcal{T} u + 2 i \alpha a_l e^{i \beta y}, & x = -D, \\
\dfrac{\partial u}{\partial x} =  \mathcal{T} u - 2 i \alpha a_r e^{i \beta y}, & x = D.
 \end{array} \right.
\end{equation}
In the $y$-direction, $u$ satisfies the quasi-periodic conditions
\begin{eqnarray}
\label{eq:BCquasi} && u(x,L/2) = e^{i \beta L} u(x,-L/2), \\
&& \dfrac{\partial u}{\partial y} (x, L/2) = e^{i \beta L} \dfrac{\partial u}{\partial y} (x, -L/2).
 \end{eqnarray}
 Thus the diffraction problem is a boundary value problem of Helmholtz equation (\ref{eq:helm}) with boundary conditions Eqs.~(\ref{eq:BCxmpD}) and (\ref{eq:BCquasi}).  Function $u_L$ ($u_R$) is a solution of the boundary value problem with $a_l =1 $ and $a_r = 0$ ($a_l=0$ and $a_r=1$).

To derive Eq.~(\ref{eq:k1_left}), we let $\tilde{u}_L$ be the diffraction solution of the  perturbed structure $\tilde{\epsilon}$ at zero-reflection frequency $\tilde{\omega}_L$ (i.e. $\tilde{k}_L = \tilde{\omega}_L/c$) for a left  incident wave with unit amplitude.  Then $\tilde{u}_L$ is a solution of the boundary value problem with $k_0$ replaced by $\tilde{k}_L$, $a_l=1$ and $a_r = 0.$ We expand  $\tilde{u}_L$, $\tilde{\mathcal{T}}$ and $\tilde{\alpha}_j$ for $\tilde{k}_L$ in power series of $\delta$:
\begin{eqnarray}
\label{eq:exp_uL} \tilde{u}_L & = & u_L + u_{1} \delta + u_{2} \delta^2 + \ldots, \\
\label{eq:exp_T} \tilde{\mathcal{T}} &= &\mathcal{T} + \mathcal{T}_1 \delta + \mathcal{T}_2 \delta^2 + \ldots,\\
\label{eq:exp_alpha} \tilde{\alpha}_j &=& \sqrt{\tilde{k}_L^2 - \beta_j^2} = \alpha_j + \gamma_{1j} \delta + \gamma_{2j} \delta^2 + \ldots,
\end{eqnarray}
for $j = 0, \pm1, \pm2, \ldots.$ Similar to the definition of $\mathcal{T}$ in Eq.~(\ref{eq:T}), the actions of $\tilde{\mathcal{T}}$, $\mathcal{T}_1$ and $\mathcal{T}_2$ on $e^{i \beta_j y}$ are simply $e^{i \beta_j y}$ multiplied by $i \tilde{\alpha}_j$, $i \gamma_{1j}$ and $i \gamma_{2j}$, respectively. 
Using expansion Eq.~(\ref{eq:exp_freq1}), we have
\begin{equation}
\label{eq:gamma}
 \gamma_{1j} = \dfrac{  k_0 k_{1}}{\alpha_j}, \quad \mbox{for} \quad j = 0, \pm1, \pm2, \ldots.
 \end{equation}

Substituting expansions (\ref{eq:eps_perturbation}), (\ref{eq:exp_freq1})  and (\ref{eq:exp_uL})-(\ref{eq:exp_alpha}) into the boundary value problem for $\tilde{u}_L$, and comparing the coefficient of $\delta$, we have
\begin{equation}
\label{eq:equ_uL1} \left\{ \begin{array}{ll} 
\Delta u_{1} + k_0^2 \epsilon u_{1} = - (2 k_0 k_1 \epsilon +  k_0^2 F) u_L, & (x,y) \in \Omega_D \\
\dfrac{\partial u_{1}}{\partial x} = - \left( \mathcal{T} u_{1} + \mathcal{T}_1 u_L \right) + 2 i \gamma_{10} e^{i \beta y}, & x = -D, \\
\dfrac{\partial u_{1}}{\partial x} =   \mathcal{T} u_{1} + \mathcal{T}_1 u_L , & x = D,
  \end{array} \right.
\end{equation}
and $u_{1}$ satisfies the quasi-periodic condition Eq.~(\ref{eq:BCquasi}) in the $y$ direction. Let 
\begin{eqnarray*}
&& \label{eq:tilde_u_xpmD_fourier}  \tilde{u}_L(- D,y) =  e^{i \beta y} + \sum\limits_{j=-\infty}^{\infty} \tilde{b}^{-}_j e^{i \beta_j y},  \\
&& \tilde{u}_L( D,y) =   \sum\limits_{j=-\infty}^{\infty} \tilde{b}^{+}_j e^{i \beta_j y},
\end{eqnarray*} 
where $\{ \tilde{b}^{+}_j \} $ are the Fourier coefficients of $\tilde{u}( D, y)$ and $\{ \tilde{b}^{-}_j \}$ are the Fourier coefficients of $\tilde{u}( - D, y) - e^{i \beta y}$, then
$$\tilde{b}^-_0 = 0, \quad  \left| \tilde{b}^+_0 \right| = 1.$$
Let 
$$ u_1(\pm D,y) = \sum\limits_{j=-\infty}^{\infty} c^{\pm}_j e^{i \beta_j y}, $$
where $\{ c_j^{\pm} \}$ are the Fourier coefficients of $u_1(\pm D, y)$, then we must have
$c^-_0 = 0. $

From the governing equations of $u_R$ and $u_1$, we have
\begin{eqnarray*} &&  - (2 k_0 k_1 \epsilon +  k_0^2 F) u_L \bar{u}_R  = \bar{u}_R (\Delta u_1 + k^2_0 \epsilon u_1)  \\
&& - u_1 (\Delta \bar{u}_R + k_0^2 \epsilon \bar{u}_R) = \nabla \cdot (\bar{u}_R \nabla u_1) - \nabla \cdot ({u_1} \nabla \bar{u}_R).
\end{eqnarray*}
Integrating the above equation on domain $\Omega_D$, we obtain
\begin{eqnarray}
\label{eq:k1} && \int_{\partial \Omega_D}  \left( \bar{u}_R \dfrac{\partial u_1}{\partial \nu} - {u_1} \dfrac{\partial \bar{u}_R}{\partial \nu}    \right) ds  \nonumber \\ 
&& = - \int_{\Omega_D} (2 k_0 k_1 \epsilon + k_0^2 F) u_L \bar{u}_R dxdy.
 \end{eqnarray}
Due to the quasi-periodic condition (\ref{eq:BCquasi}), the line integrals on the two edges of $\Omega_D$ at $y = \pm L/2$  cancel out. Furthermore, by using the boundary conditions of $u_1$ at $x = \pm D$ and expansions of $u_L, u_R$ and $u_1$ at $x = \pm D$, the left-hand side of Eq.~(\ref{eq:k1}) can be reduced to 
$$ i L \gamma_{10} (b^+_0 - \bar{b}^+_0) + i L \sum\limits_{\substack{j=-\infty \\ j \neq 0}}^{\infty}  \gamma_{1j} ( \bar{b}^-_j b^+_j + \bar{b}^-_j b^+_j ). $$
In the above, the conditions $b_0^- = c_0^- = 0$ are used.
Substituting the above into Eq.~(\ref{eq:k1}) and noticing the formula of $\gamma_{1j}$ (i.e. Eq.~(\ref{eq:gamma})), we obtain  Eq.~(\ref{eq:k1_left}).  Equation~(\ref{eq:k1_right}) can be  similarly derived.







\end{document}